\documentclass[12pt]{iopart}
\usepackage{graphicx}

\newcommand{\eg}{e.g.}
\newcommand{\msun}{\ensuremath{M_{\odot}}}

\newcommand{\nuc}[2]{\ensuremath{\mathrm {^{#2}#1}}}
\newcommand{\gcc}{\ensuremath{\mathrm{\thinspace g \thinspace cm^{-3}}}}

\begin{document}

\jl{4}

\title{Supernova Science at Spallation Neutron Sources}

\author{W Raphael Hix\dag\ddag\S, Anthony Mezzacappa\dag, O E Bronson Messer\ddag\dag\S and Stephen W Bruenn$\parallel$}

\address{\dag\ Physics Division, Oak Ridge National Laboratory, Oak Ridge, TN 37831-6354 USA} 

\address{\ddag\ Department of Physics and Astronomy, University of Tennessee, Knoxville, TN 37996-1200 USA}

\address{\S\ Joint Institute for Heavy Ion Research, Oak Ridge National Laboratory, Oak Ridge, Tennessee 37831-6374 USA}

\address{$\parallel$ Department of Physics, Florida Atlantic University, Boca Raton, FL 33431 USA}

\begin{abstract}
New facilities to measure neutrino-nucleus cross sections, such as those possible in conjunction with spallation neutron sources, could provide an experimental foundation for the many neutrino-nucleus weak interaction rates needed in supernova models. This would enable more realistic supernova models and provide a greatly improved ability to understand the physics fundamental to supernovae by comparison of these models with detailed observations.  Charged- and neutral-current neutrino interactions on nuclei in the stellar core play a central role in supernova dynamics and  nucleosynthesis as well as being important for supernova neutrino detection. Measurements of these reactions on judiciously chosen targets  would provide an invaluable test of the complex theoretical models used to compute the large number of neutrino-nucleus cross sections that are needed.
\end{abstract}


\section{Core Collapse Supernovae}

Core collapse supernovae are among the most energetic explosions in the 
Universe, releasing $10^{46}$ Joules of energy in the form of neutrinos of 
all flavors at the staggering rate of $10^{57}$ neutrinos per second and 
$10^{45}$ Watts.  Marking the death of a massive star (mass $> 8-10$ solar masses) and the birth of a neutron star or black hole, core collapse supernovae are a nexus of nuclear physics, particle physics, fluid dynamics, radiation transport, and general relativity. They serve as laboratories for physics beyond the Standard Model and for matter at extremes of density, temperature, and neutronization 
that cannot be produced in terrestrial laboratories.  The $10^{44}$ Joules of kinetic energy and rich mix of recently synthesized elements delivered into the interstellar medium by the ejecta of each supernova make core collapse supernovae a key link in our chain of origins from the Big Bang to the formation of life on Earth.

The center of a massive star as it nears its demise is composed of iron,
nickel, and similar elements, the end products of stellar nucleosynthesis. 
Above this \emph{iron core} lie concentric layers of successively lighter
elements, recapitulating the sequence of nuclear burning that occurred in
the core during the star's lifetime.  Unlike prior burning stages, where the ash of one stage became
the fuel for its successor, no additional nuclear energy can be released by
further fusion of the maximally bound, iron peak nuclei.  No longer can nuclear energy production stave off the inexorable force of gravity.  When the iron core grows too massive to be supported by electron degeneracy pressure, the core collapses.  This collapse continues until the core reaches densities similar to those of the nucleons in a nucleus, whereupon the repulsive core of the nuclear interaction renders the core incompressible, halting the collapse.  Collision of the supersonically falling overlying layers with this stiffened core produces the \emph{bounce shock}, which drives these layers outward.  However, this bounce shock is sapped of energy by the escape of neutrinos and nuclear dissociation and stalls before it can drive off the envelope of the star (see, \eg, \cite{BuLa85}).  The failure of this \emph{prompt} supernova mechanism sets the stage for a \emph{delayed} mechanism, wherein the intense neutrino flux, which is carrying off the binding energy of the proto-neutron star, heats matter above the neutrinospheres and reenergizes the shock \cite{Wils85,BeWi85}.  The heating is mediated primarily by the  absorption of electron neutrinos and antineutrinos on the dissociation-liberated free nucleons behind the shock.  This process depends critically on the neutrino luminosities, spectra, and isotropy, i.e., on the multigroup (multi-neutrino energy) and multi-angle neutrino transport between the proto-neutron star and the shock.

Realistic supernova models will require extremely accurate neutrino radiation hydrodynamics, but, as we will show, unless advancements in neutrino transport and hydrodynamics are matched by equally important advancements in nuclear and weak-interaction physics, the accuracy of the former must be called into question.  For electron and electron-neutrino capture rates, and in general all neutrino-nucleus cross sections, detailed shell model computations must replace the parameterized approximations that have been used in the past. For example, recent work \cite{LaMa00} on electron capture up to atomic mass 65 has shown that these parameterized rates can be orders of magnitude in error. The electron capture rate is dominated by Gamow-Teller resonance transitions, with the Gamow-Teller strength distributed over 
many states. Previous parameterized treatments \cite{FuFN85} place the resonance at a single
energy, and this energy is often grossly over- or underestimated
relative to the Gamow-Teller centroid computed in realistic shell 
model computations, leading in turn to rates that are often 
too small or too large, respectively. (If the centroid is underestimated, more electrons can participate in capture.  The reverse is true if the centroid is too high.)   To take full advantage of these improved rates, improvements in the tracking of the nuclear composition will be needed.  The use of a single representative nucleus, which has been the standard until now in virtually all supernova models \cite{BaCK85,LaSw91}, can underestimate the rates of electron and electron-neutrino capture during the critical core collapse phase (see, \eg, \cite{AFWH94}). 

While generations of nuclear structure models will afford 
ever greater realism in the calculation of stellar core properties
and the interactions of nuclei with the neutrinos flowing through 
the core, nuclear experiments must be designed and carried out that will serve as guide posts for the theoretical predictions that are required to produce the large number of rates that enter into any realistic supernova 
or supernova nucleosynthesis model. In particular, we must have 
neutrino-nucleus cross section measurements that will help gauge 
the accuracy of neutrino capture and scattering predictions important during stellar core collapse and during $\alpha$-rich freezeout and the r-, p- and $\nu$ nucleosynthesis processes after core bounce.

\section{Spherically Symmetric Supernova Models}

Although four decades of supernova modeling have established the textbook explanation discussed in the previous section, fundamental questions about the explosion mechanism remain.  Is the neutrino heating sufficient, or are multidimensional effects such as convection and rotation necessary?  Can the basic supernova observable, an explosion, be reproduced by detailed spherically symmetric models, or are multidimensional models required?   Many observations of core collapse supernovae show clear violations of spherically symmetry. For example, neutron star kicks \cite{FrBB98} and the polarization of supernova emitted light \cite{WHHW01} cannot arise in spherical symmetry.  Nonetheless, it remains uncertain whether such observations point to fundamental features of the supernova mechanism or merely side effects. To answer this question, simulations in one, two, and three dimensions must be coordinated.

The neutrino energy deposition behind the shock depends sensitively 
on the neutrino luminosities, spectra, and angular distributions in 
the postshock region. Ten percent variations in any of these quantities 
can make the difference between explosion and failure in supernova 
models \cite{JaMu96,BuGo93}. Thus, accurate multigroup Boltzmann neutrino 
transport must be considered in supernova models. Past spherically symmetric simulations have implemented increasingly  sophisticated approximations to Boltzmann transport: simple leakage  schemes \cite{VaLa81}, two-fluid models \cite{CoVB86}, and multigroup  flux-limited diffusion \cite{Arne77,Brue85,MBHL87,WiMa93}.  A generic feature of this last, most sophisticated approximation is that it underestimates 
the isotropy of the neutrino angular distributions in the heating region 
and, thus, the heating rate \cite{Jank92,MMBG98a}.  With these limited transport approximations came the possibility that the failure to produce explosions in the past may have resulted from the incomplete neutrino transport.

To address this possibility, the ORNL theoretical astrophysics group has used Boltzmann neutrino transport to successfully simulate the core collapse, bounce and evolution for more than 500 milliseconds after bounce of a number of spherically symmetric models.  Even with our implementation of complete Boltzmann neutrino transport, via AGILE-BOLTZTRAN, models in both the Newtonian \cite{MLMH01} and the general relativistic \cite{LMTM01} limits have failed to produce explosions for a range of progenitor masses from 13-40 solar masses \cite{NoHa88,WoWe95,HLMW01}.  These results are supported by the failure of a 15 solar mass model to explode using a different implementation of Boltzmann transport \cite{RaJa00}.  In all cases, neutrino heating propels the shock to a radius of more than 150 km; however, the heating is insufficient to eject the envelope, so the shock recedes.  In the general relativistic model starting from a 40 \msun\ progenitor \cite{WoWe95}, the formation of a black hole is evident approximately one half second after bounce.
 
\begin{figure}[tp]
  \begin{center}
    \includegraphics[width=\textwidth]{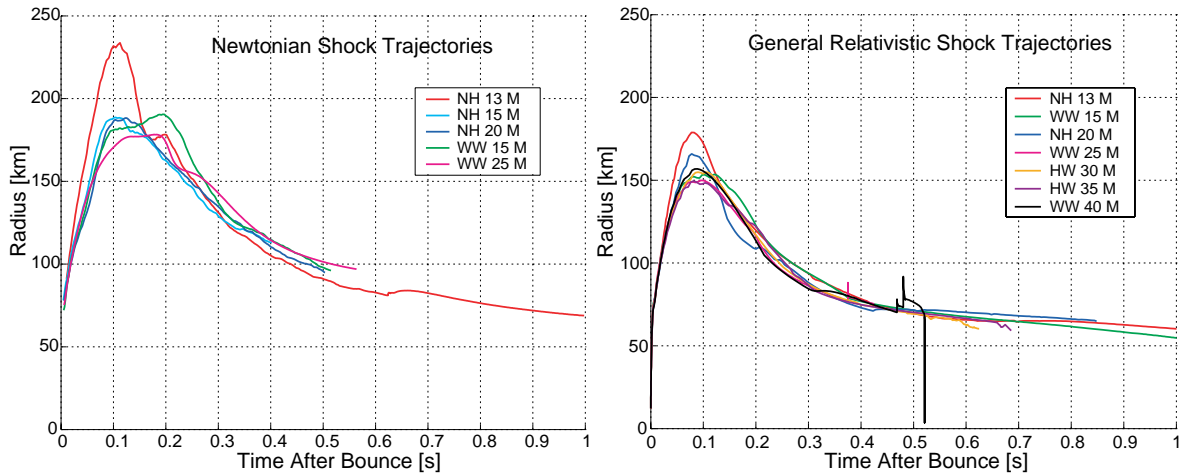} 
  \end{center}
  \caption{We trace the shock fronts for Newtonian (left) and General Relativistic (right) supernova models which fail to produce explosions for a range of progenitors \cite{NoHa88,WoWe95,HLMW01}}
\end{figure}

Thus, we are beginning to answer some fundamental questions in supernova 
theory. In the past, it was not clear whether failure or success in supernova models was the result of inadequate transport approximations.  However, because our failed explosions were obtained using accurate spherically symmetric hydrodynamics \cite{LiMT01}, state of the art neutrino transport \cite{MeMe98}, industry standard neutrino-matter interactions \cite{Brue85} and the equation of state of \cite{LaSw91}, our conclusion is unambiguous.  Accurate neutrino transport alone does not overcome the failure of supernova simulations that assume spherical symmetry to produce explosions.  This would suggest that changes in our 
input physics (i.e., weak interaction physics and nuclear physics in the Equation of State) and/or initial conditions (i.e., stellar evolution models) are needed and/or that multidimensional effects such as convection, rotation, and magnetic fields are required ingredients in the recipe for explosion.  

\section{Neutrino-Nucleus Interactions and Core Collapse Dynamics}

\begin{figure}[t]
\begin{center}
\includegraphics[width=\textwidth]{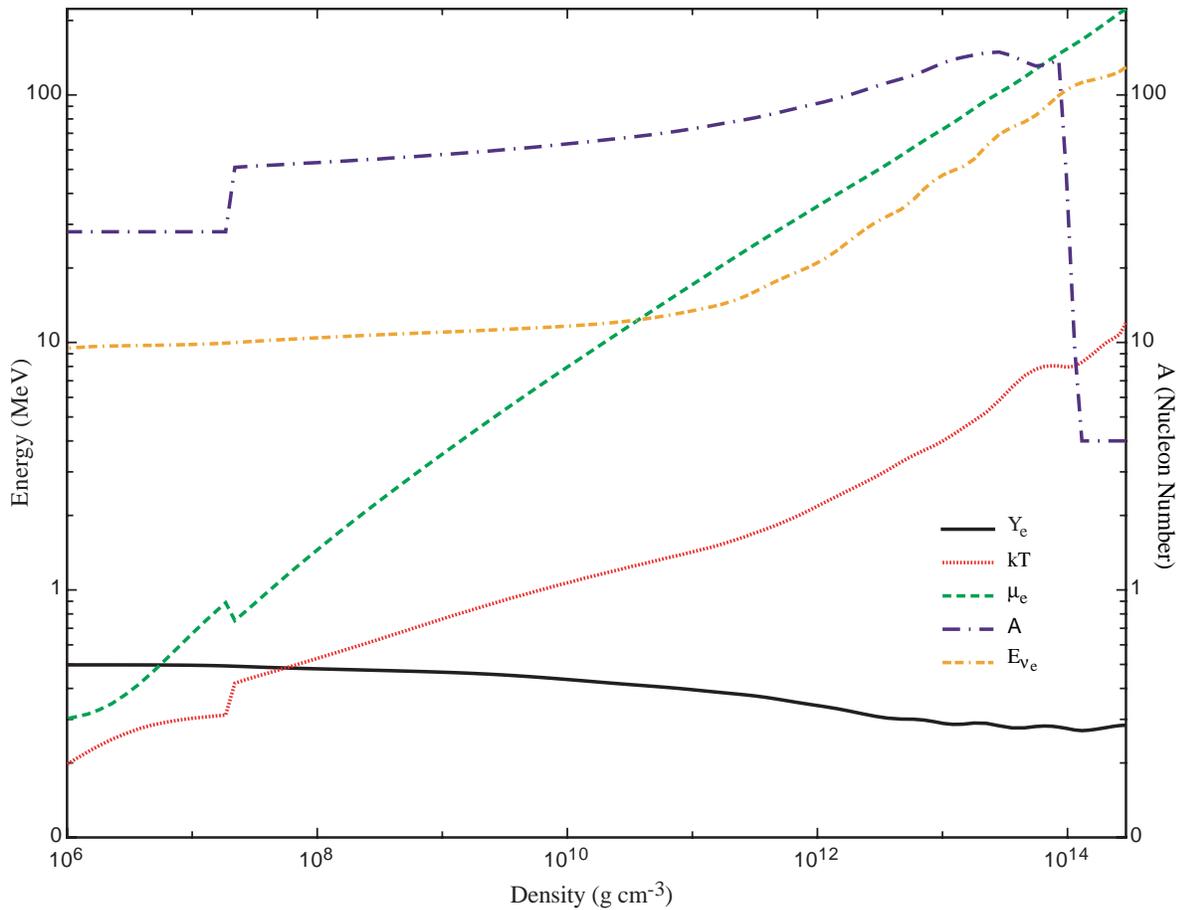}
\caption{The energy scales and composition as a function of density in the collapsed stellar core at bounce for a 15 \msun\ progenitor \cite{HLMW01}.}
\label{fig:corethermo}
\end{center}
\end{figure}

\begin{figure}[t]
\begin{center}
\includegraphics[angle=270,width=.8\textwidth]{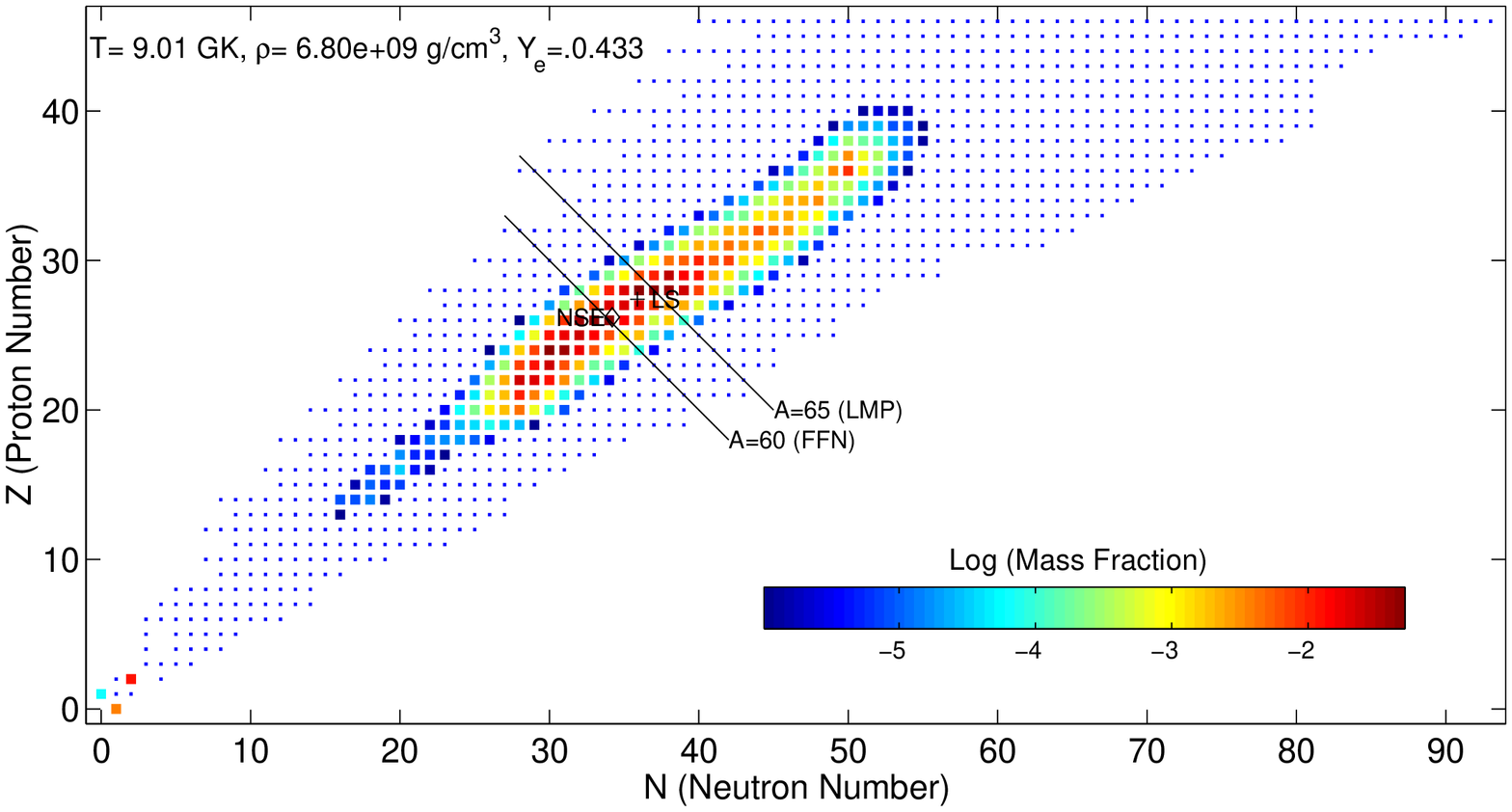}

\includegraphics[angle=270,width=.8\textwidth]{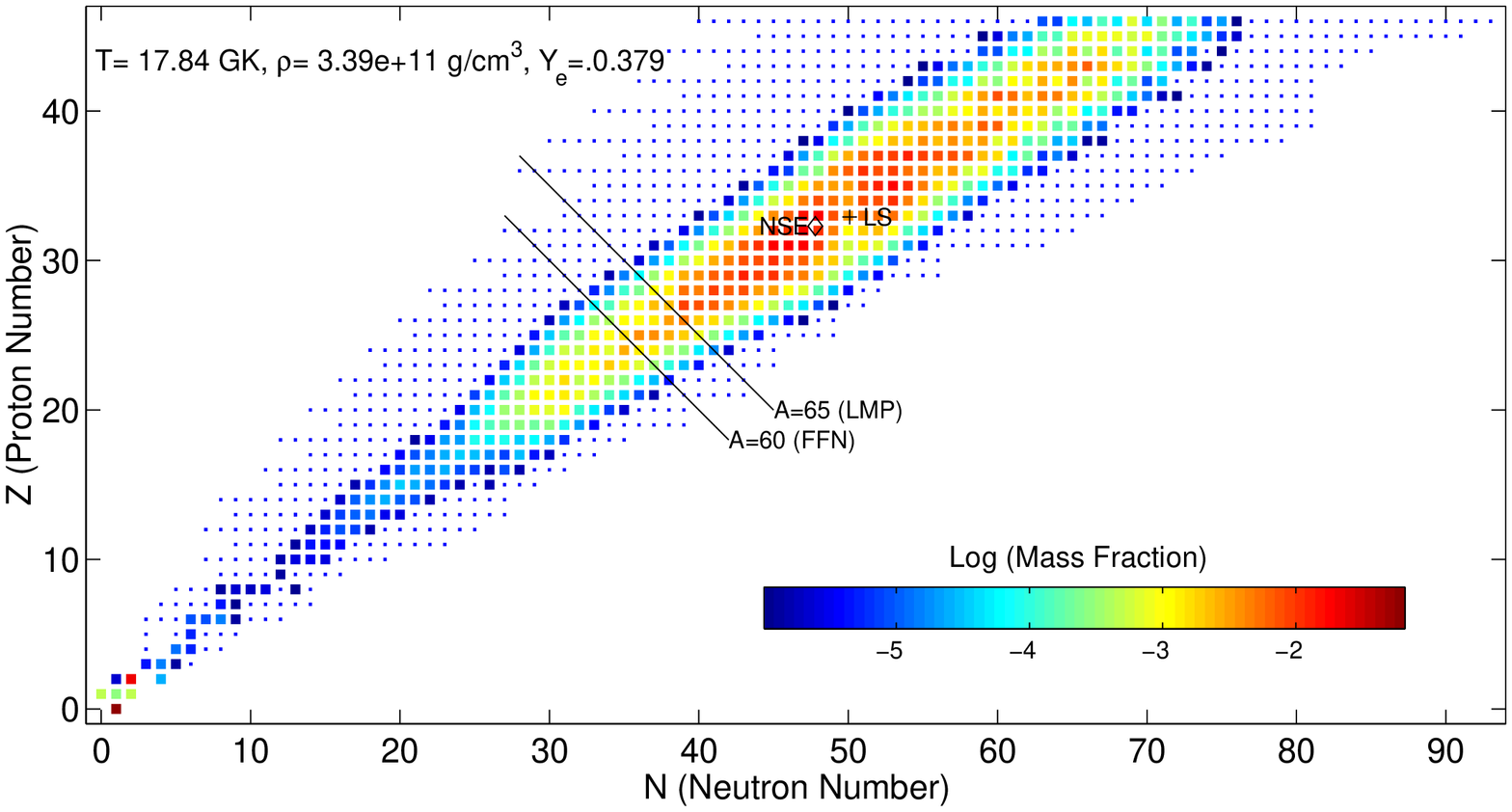}
\caption{Details of the composition at two points during stellar core collapse \cite{LMMH01}.}
\label{fig:corecomp}
\end{center}
\end{figure}

While neutrino interactions with shock heated nucleons are the major source of the neutrino heating which drives the delayed shock, neutrino interactions with nuclei are also important before and after the passage of the shock.  During the collapse of the stellar core, electrons are captured on heavy nuclei and free protons in the core, producing electron neutrinos that initially escape, \emph{deleptonizing} the core.   Where the shock forms in the stellar core at bounce and how much energy it initially has are set by the amount of deleptonization in the core during collapse.  Deleptonization would be complete if electron capture continued without competition, but at densities of order $10^{11-12} \gcc$, the electron neutrinos become ``trapped'' in the core, and the inverse  reactions---charged-current electron-neutrino capture on iron-peak nuclei and free neutrons---begin to compete with electron capture until the reactions are in weak equilibrium and the net deleptonization of the core ceases on the core collapse time scale. The equilibration of electron neutrinos with the stellar core occurs at densities between $10^{12-13} \gcc$.  Figure~\ref{fig:corethermo} summarizes the thermodynamic conditions throughout the core at bounce and displays the temperature, electron fraction ($Y_e$), electron chemical potential ($\mu_e$), and mean electron neutrino energy ($E_{\nu_e}$) in MeV as functions of the matter density.  Also shown is the representative nuclear mass ($A$).  The kinks near $3 \times 10^{7} \gcc$ mark the transition to the silicon shell.  As the stellar core densities increase, the characteristic nuclei in the core increase in mass, owing to a competition between Coulomb contributions to the nuclear free energy and nuclear surface tension.  For densities of order $10^{13} \gcc$, nuclei with mass $>100$ dominate.  For densities exceeding $10^{14} \gcc$, heavy nuclei are replaced by nuclear matter.  Figure~\ref{fig:corecomp} demonstrates that the nuclear composition shows a wide spread in mass, with species with significant concentrations having masses that differ by 40.   Further, Figure~\ref{fig:corecomp} also shows that the abundances of nuclei with mass  greater than 100 are significant as early as $10^{11} \gcc$.  Thus, cross sections for charged-current electron and electron-neutrino capture on nuclei at least through mass 120 are needed to accurately simulate core deleptonization and to accurately determine the postbounce initial conditions. 

Improved shell model calculations of weak interaction rates for electron capture, positron capture, and $\beta^{-}$ and $\beta^{+}$ decays on nuclei relevant for stellar evolution $(45<A<65)$ have become available in recent years \cite{LaMa00,OHMT94}.  Heger \etal\ \cite{HLMW01} have utilized these new weak reaction rates to improve upon the stellar evolution simulations of Woosley \& Weaver (WW95)\cite{WoWe95}, replacing the weak interaction rates for electron and positron captures and $\beta^{-}$ and $\beta^{+}$ decays.  The WW95 models used the electron capture rates of Fuller, Fowler, \& Newman (FFN) \cite{FuFN85} and older sets of beta decay rates \cite{Mazu73,Hans66}.  The most noticeable effect of these improvements is a
marked increase in the electron fraction ($Y_{e}$) throughout the iron core before collapse. Because the final size of the homologous core, and therefore the shock formation radius, is proportional to the square of the trapped lepton fraction (${Y_{l}}^{2}$) at core bounce \cite{Yahi83}, the persistence of these initial differences in $Y_{e}$ throughout collapse could have a discernible effect
on the shock energetics.

In an initial attempt to determine the influence of these improved stellar evolution models on iron core collapse, Messer \etal\ \cite{MLMH03} performed parameterized radiation hydrodynamics collapse simulations using AGLE-BOLTZTRAN \cite{MeMe98, MeBr93a, LiRT02, LMMB03}.  These simulations show no difference in initial shock formation position between the two sets of progenitor models when ``standard'' weak interaction physics is used during core collapse.  Unlike stellar evolution simulations, the only available information about the nuclear composition in supernova simulations is that provided by the nuclear equation of state \cite{LaSw91}: mass fractions of free nucleons, $\alpha$-particles, and heavy nuclei, as well as the mean charge (Z) and atomic mass (A) of the heavy nuclei.  This limited information precludes the use of reaction rates calculated for individual isotopes, which were, in any event, unavailable for $A>65$.  Figure~\ref{fig:corecomp} depicts the limits of the FFN and LMP data sets, demonstrating their usefulness for stellar evolution, but their inadequacy for core collapse.  Instead, the Messer \etal\ simulations used the standard neutrino emissivity from nuclei developed by Bruenn \cite{Brue85} (see also \cite{MeBr93b}).  This prescription, based on the independant particle model, treats electron capture on heavy nuclei through a generic $0f_{7/2} \rightarrow 0f_{5/2}$ Gamow-Teller transition in the average heavy nucleus identified by the equation of state.  Because this treatment does not include additional Gamow-Teller transitions, forbidden transitions, or thermal unblocking \cite{Full82,CoWa84,LaKD01}, electron capture on heavy nuclei ceases when the neutron number of the average nucleus exceeds 40.  As a result, electron capture on protons dominates the later phases of collapse.  The steep dependence of the free proton fraction on changes in the electron fraction assures convergence of initially distinct $Y_{e}$ profiles to a similar $Y_{e}$ profile inside the homologous core and, consequently, to a similar shock formation radius.  However, when the parameter in the models that controls the rate of electron capture on nuclei is adjusted so that the capture on protons does not dominate, significant changes in the core collapse dynamics are evident. This demonstrates the need for accurate treatment of nuclear electron capture during stellar core collapse. 

Recently, Langanke \etal\ \cite{LMSD03} have produced electron capture rates for a sample of nuclei with $A=66-112$ using hybrid, shell-model--RPA calculations.  As a major advance over the simple treatment of nuclear electron capture in previous supernova simulations, Hix \etal \cite{HMML03} have developed a treatment of nuclear electron capture based on these hybrid rates and the shell model electron capture rates from Langanke \etal \cite{LaMa00} (LMP) for $A\leq65$.  For the distribution of emitted neutrinos, the approximation described by Langanke \etal\ \cite{LaMS01} was used.  To calculate the needed abundances of the heavy nuclei, a Saha-like NSE is used \cite{ClTa65,HaWE85}, including Coulomb corrections to the nuclear binding energy \cite{HixPhD,BrGa99}, but neglecting the effects of
degenerate nucleons \cite{ElHi80}.  This NSE treatment has been used in
prior investigations of electron capture in thermonuclear supernovae
\cite{BDHI00}.  The combined set of LMP and hybrid model rates are used to
calculate an average neutrino emissivity per nucleus.  The full neutrino
emissivity is then the product of this average with the number density
of heavy nuclei calculated by the equation of state.  With the limited
coverage of rates for $A>65$, this approach provided the most reasonable
estimate of what the total electron capture would be if rates for all
nuclei were available.  This averaging approach also makes the rate of
electron capture consistent with the composition returned by the EOS,
while minimizing the impact of the limitations of the NSE treatment.   

\begin{figure}
  \begin{center}
  \includegraphics[height=4in,width=.8\textwidth]{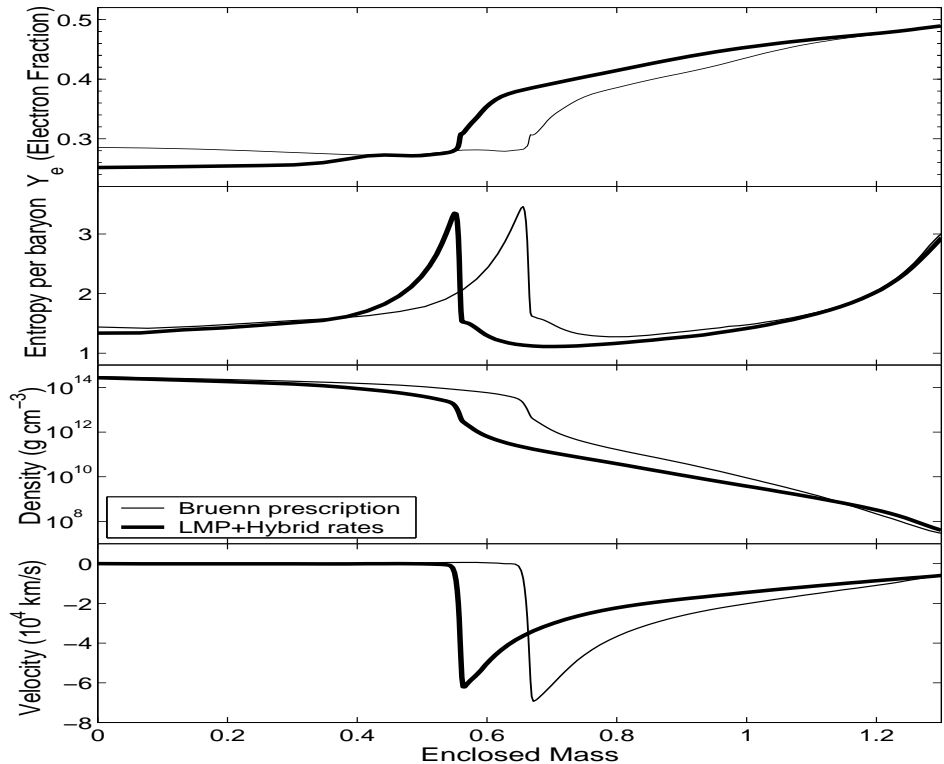}
  \caption{The electron fraction, entropy, density and velocity as functions 
  of the enclosed mass at the beginning of bounce for a 15 \msun\ model.  
  The thin line is a simulation using the Bruenn parameterization while the 
  thick line is for a simulation using the LMP and hybrid reaction rate sets.
  \label{fig:bounce}}
  \end{center}
\end{figure}

This improved treatment of nuclear electron capture has two competing
effects.  In lower density regions, where the average nucleus is well
below the $N=40$ cutoff of electron capture on heavy nuclei, the Bruenn
parameterization results in more electron capture than the LMP+hybrid
case.  This is similar to the reduction in the amount of electron
capture seen in stellar evolution models \cite{HLMW01} and thermonuclear
supernova \cite{BDHI00} models when the FFN rates are replaced by shell
model calculations.  In denser regions, the continuation of electron
capture on heavy nuclei alongside electron capture on protons results in
more electron capture in the LMP+hybrid case.  The results of these
competing effects can be seen in the upper pane of Figure~\ref{fig:bounce}, which shows the distributions of $Y_e$ throughout the core at bounce.  In addition to the marked reduction (11\%) in the electron fraction in the interior of the PNS, this improved treatment also results in a nearly 20\% reduction in the mass of the homologous core, which manifests itself as a reduction in the mass interior to the formation of the shock from .67 \msun\ to .57 \msun\ in the LMP+hybrid case.  A shift of this size is very significant dynamically because the dissociation of .1 \msun\ of heavy nuclei by the shock costs $10^{51}$ erg, the equivalent of the explosion energy.  There is also an 11\% reduction in the central density and a 7\% reduction in the central entropy at bounce, as well as a 10\% smaller velocity difference across the shock and quite different lepton and entropy gradients throughout the core.  In the outer regions, the higher electron fraction slows collapse, resulting in, for example, reductions of a factor of 5 in density and 40\% in velocity in the vicinity of 0.8 \msun.

These differences in the behavior of collapsing stellar cores illustrates the importance of weak interactions with nuclei.  At the onset of collapse, the nuclei of interest are clustered in mass between 50 and 70 along the neutron-rich edge of stability. Throughout collapse, decreasing electron fraction and increasing density pushes the composition to heavier and more neutron-rich nuclei, including nuclei 4-6 decays away from stability and with masses greater than 100.  The KARMEN collaboration pioneered work in this regime, measuring the cross section for $\nuc{Fe}{56}(\nu_{\rm e},e^{-})\nuc{Co}{56}$ \cite{Masc98}, which is one of the nuclei of interest early in collapse.  However, the sheer number of potentially important species, and their instability,makes direct measurements of the rates an impossibility.  Nonetheless, measurements of the relevant neutrino-nucleus interactions remain extremely valuable by providing the most relevant constraints on the theoretical models.  The technique used by the KARMEN collaboration can be used to measure the electron-neutrino capture cross section on any of a wide range of nuclei, wherein the natural abundance is dominated by a single isotope and the element is a solid at room temperature.  Several such nuclei are in the critical nuclear mass range:  \nuc{Mn}{55}, \nuc{Co}{59}, \nuc{Y}{89}, \nuc{Nb}{93}, \nuc{Rh}{103} and \nuc{In}{115}.  Priority among these choices should be decided by their ability to constrain the theoretical rate calculations.

\section{Breaking Spherical Symmetry}

Multi-dimensional simulations allow investigation of the role convection, rotation, and magnetic fields may play in the explosion.  Supernova fluid instabilities fall into two categories: (1) instabilities near or below the neutrinospheres, which we refer to as proto-neutron star instabilities and (2) convection between the gain radius and the shock, which we refer to as neutrino-driven convection.  Proto-neutron star instabilities may aid the explosion mechanism by transporting hot, lepton-rich matter to the neutrinospheres, thereby boosting the luminosities at the neutrinosphere. Neutrino-driven convection may aid the explosion mechanism by boosting the shock radius and the neutrino heating efficiency, thereby facilitating shock revival.  While a presentation of these macroscopic phenomena may seem out of place in a discussion of the role of microscopic neutrino-nucleus interactions, as we will demonstrate, the linkage of the microscopic and macroscopic scales are part of what make supernovae both interesting and challenging to study.

\subsection{Proto-Neutron Star Instabilities}

\begin{figure}
  \begin{center}
  \includegraphics[height=4in,width=.8\textwidth]{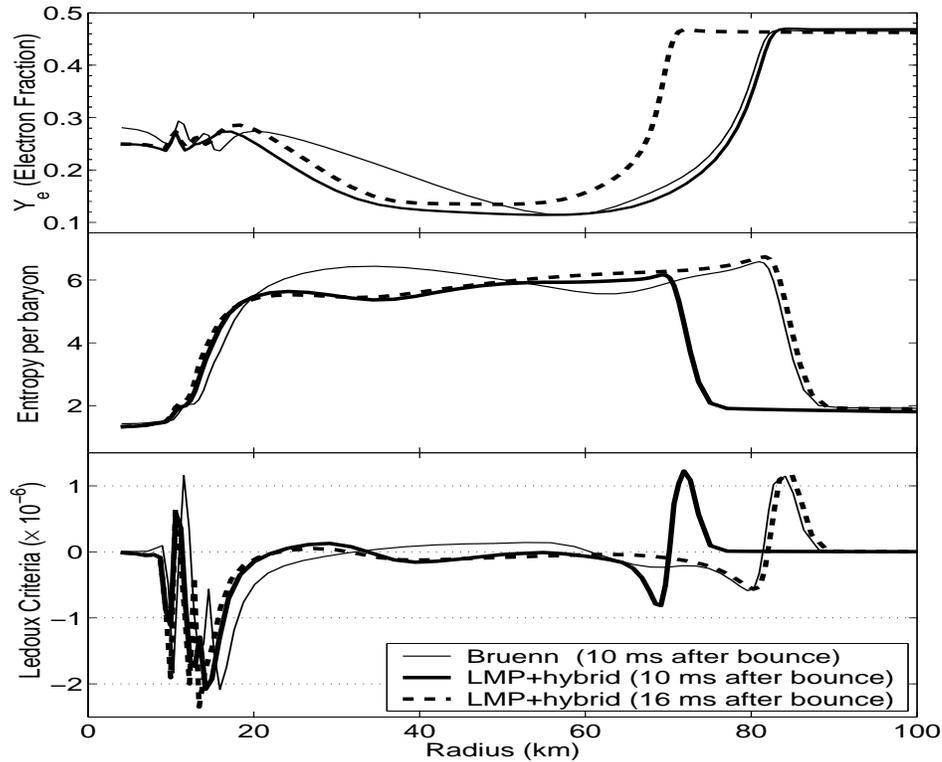}
  \end{center}
  \caption{The electron fraction, entropy and Ledoux criterion as functions   
  of the radius soon after bounce.  The thin line (B10) shows results a 
   simulation using the Bruenn parameterization at 10 milliseconds after bounce while the thick lines are from the LMP+hybrid simulation set at 10 (solid, LMPH10) and 16 (dashed, LMPH16) milliseconds after bounce. \label{fig:ledoux}}
\end{figure}

Within the proto-neutron star (PNS), whose ``surface'' is defined by the neutrinospheres, a number of fluid instabilities may arise as the result of the lepton fraction and entropy gradients present.  The lepton gradients are established by the deleptonization of the proto-neutron star via electron-neutrino escape near the electron neutrinosphere, whereas the entropy gradients result from the weakening supernova shock. (As the shock weakens, it causes a smaller entropy jump in the material flowing through it.)  It is by its influence on these gradients that the microscopic physics links to the macroscopic fluid behavior.  As an example, we return to our comparison in the previous section of models using an improved description of nuclear electron capture.  Figure~\ref{fig:ledoux} shows the entropy and lepton gradients near the onset of PNS convection.  The curves labeled B10 and LMPH10 show the results of the Bruenn and LMP+hybrid cases, both 10 ms after their respective maximum densities are reached. Clearly visible in the region between 20 km and 60 km are an inward displacement and steepening (flattening) of the negative lepton (entropy) gradient that results from our more accurate treatment of electron capture.  The curves labeled LMPH16 present the LMP+hybrid model at 16 ms after bounce, when its shock has reached a radius matching that of B10.  Clearly the differences in the gradients described above are not transient, nor do they arise from the slower progress of the shock in the LMP+hybrid case.

To illustrate the effects these changes in the entropy and lepton
gradients could have on convection in the proto-neutron star, we have plotted the Ledoux criterion $(\partial \ln \rho/\partial \ln Y_{e}) (d \ln Y_{e} / dr) + (\partial \ln \rho/\partial \ln s) (d \ln s/dr)$ in the lower pane of Figure~\ref{fig:ledoux}.  Instability in the Ledoux approximation arises when this expression is positive.  The broad spikes on the right that
coincide with the shock are due to the numerical spreading of the shock and, thus, are an artifact.  In the model employing the Bruenn treatment, there is an extended region of instability between 30 and 60 km, while in the LMP+hybrid case, the unstable region is only 10 km across and displaced inward.  This suggests that, with an accurate treatment of nuclear electron capture, PNS convection would be less extensive than previously thought and would occur deeper within the stellar core, though the similarity of the values of the Ledoux criterion suggests that it would occur with similar strength.

While this spherically symmetric example demonstrates the importance of neutrino interactions in determining the dynamic behavior of the fluid, it is far from the final answer.  The development of convection in the proto-neutron star is a radiation hydrodynamic phenomenon, rather than a purely hydrodynamic phenomenon. In this region of the stellar core, neutrinos and the stellar core fluid are coupled. The neutrinos have the ability to equilibrate an otherwise buoyant fluid element with its surroundings in both lepton number and entropy, rendering the fluid element nonbuoyant \cite{mcbb98a}. Thus, a determination of the degree of Ledoux convection in this region, and its extent, must await three-dimensional radiation hydrodynamics simulations. Past studies, in one respect or another, could not fully determine either. In one case \cite{mcbb98a}, neutrino transport was assumed to be spherically symmetric, which overestimates the degree to which the fluid element and its surroundings may equilibrate, and excluded the innermost 25 km of the core, in which it is unlikely that equilibration will occur owing to the large neutrino interaction cross sections and inefficient transport. On the other hand, other detailed studies \cite{KeJM96,BRJK03} did include this innermost region, but used radial-ray neutrino transport, which underestimates the degree to which equilibration will occur by disallowing lateral transport between rays. In the first study, vigorous convection did not occur, whereas in the latter, vigorous convection did occur and was confined to the inner 20 km
of the core. The full vigor and extent will be determined by fully two-dimensional, and ultimately three-dimensional, models. 

Neutron fingers are another instability that may occur in regions of crossed gradients in lepton fraction and entropy. Neutron fingers are ``doubly diffusive'' instabilities, stemming from competing efficiencies of lepton number and entropy transport in the core.   If, for example, a region of high entropy and low lepton fraction exists at a larger radius in a central gravitational field than a region of low entropy and high lepton fraction, and if entropy transport is more efficient than lepton transport, a fluid element at the interface of the two regions perturbed inward will sink, generating a low-entropy, low-lepton--fraction (neutron-rich)
finger. An early attempt to investigate these effects was made in the simulations by Wilson \& Mayle \cite{WiMa93}.  These one-dimensional models lifted spherical symmetry in an approximate fashion using a phenomenological mixing-length description for neutron finger convection inside the proto-neutron star, which boosted the neutrino luminosities, causing explosions.  However, the strength of the neutron finger instability assumed by Wilson \& Mayle is controversial, as shown by detailed numerical neutrino equilibration experiments \cite{BrDi96}. 

This range of outcomes clearly demonstrate that to determine whether 
or not proto-neutron star instabilities exist and, if they exist, are vigorous will require simulations coupling three-dimensional, multigroup neutrino transport and three-dimensional hydrodynamics.  Moreover, accurate neutrino interaction rates, for both the possibly unstable phase and the preceding collapse phase, will be essential.  

\subsection{Neutrino-Driven Convection}

Neutrino-driven convection occurs in the region behind the stalled shock but above the gain radius as a result of the entropy gradient that forms as material infalls while being continually heated from below.  Models show high-entropy, rising plumes and lower-entropy, denser, finger-like downflows beneath the shock. The shock is  ultimately distorted by this convective activity.  In the Herant \etal \cite{HBHF94} simulations, this large-scale convection led to increased neutrino energy deposition, the accumulation of mass and energy in the gain region, and a thermodynamic engine that ensured explosion. Herant \etal used two-dimensional ``gray'' (neutrino-energy-integrated, as opposed to multigroup) flux-limited diffusion in neutrino-thick regions and a neutrino lightbulb approximation in neutrino-thin regions.  In a lightbulb approximation, the neutrino luminosities and rms energies are assumed constant with radius.  Herant \etal described this transport scheme as the ``most wanting aspect'' of their simulations, stressing the need for more sophisticated multidimensional, multigroup transport in future models.  In the Burrows \etal \cite{BuHF95} simulations, neutrino-driven convection in some models significantly boosted the shock radius and led to explosions.  However, they stressed that success or failure in producing explosions was ultimately determined by the values chosen for the neutrino spectral parameters in their gray ray-by-ray (one-dimensional) neutrino diffusion scheme. (In spherical symmetry (1D), all rays are the same. In a ray-by-ray scheme in axisymmetry (2D), not all rays are the same, although the transport along each ray is a 1D problem. In the latter case, lateral transport between rays is ignored.)  Focusing on the neutrino luminosities, Janka and M\"{u}ller \cite{JaMu96}, using a central adjustable neutrino lightbulb, conducted a parameter survey and concluded that neutrino-driven convection aids explosion only in a narrow luminosity window ($\pm 10\% $), below which the luminosities are too low to power explosions and above which neutrino-driven convection is not necessary. 

In more recent simulations carried out by Swesty \cite{Swes98} using two-dimensional gray flux-limited diffusion in both neutrino-thick and neutrino-thin regions, it was demonstrated that the simulation outcome varied dramatically as the matter-neutrino ``decoupling point,'' which in turn sets the neutrino spectra  in the heating region, was varied within reasonable limits. (The fundamental difficulty with gray transport schemes is that the neutrino spectra, which are needed for the heating rate, are not computed. The spectra are specified by a neutrino ``temperature,'' normally chosen to be the matter temperature at decoupling. In a multigroup scheme, the spectra are computed differentially.)   In two-dimensional models by Mezzacappa \etal \cite{MCBB98b}, the angle-averaged shock radii do not differ significantly from the shock trajectories in  their one-dimensional counterparts, and no explosions are obtained. Neither the luminosities nor the neutrino spectra are free parameters. These two-dimensional simulations implemented spherically symmetric (1D) multigroup flux-limited diffusion neutrino transport, compromising transport spatial dimensionality to implement multigroup transport and a seamless transition between neutrino-thick and neutrino-thin regions.  Recently, Fryer and Warren \cite{FrWa02} (using methods similar to \cite{HBHF94}) have demonstrated that three dimensional models exhibit convective behavior similar to the two dimensional models.  This somewhat surprising preference for large scale convection in both two and three dimensional simulations is possibly explained by hydrodynamic instabilities in the stalled accretion shock \cite{BlMD03}.  Most recently, a simulation by \cite{BRJK03}, coupling 2D hydrodynamics with the ray-by-ray neutrino transport (performing independent calculations of the radiation transport along each radial direction) failed to produce an explosion.

In light of the neutrino transport approximations made in all multidimensional supernova simulations to date, next-generation simulations will have to reexplore neutrino-driven convection in the context of three-dimensional hydrodynamic simulations that also implement more realistic multigroup three-dimensional neutrino transport.  Furthermore, such simulations will only be as accurate as the neutrino matter interactions they include.  In the previous section, we demonstrated the significant reduction in the density and velocity of the material infalling toward the shock due to an improved prescription for nuclear electron capture (see Fig.~\ref{fig:bounce}).  In addition, charged current neutrino capture (and neutral current inelastic neutrino scattering \cite{BrHa91}) on heavy nuclei can alter the entropy and neutronization of this infalling matter prior to its arrival at the shock.  Such changes in the pre-shock matter affect the shock propagation and thermodynamic conditions in the post-shock convective region, illustrating again the dependence of the macroscopic fluid dynamics on accurate microscopic physics.

\section{Supernova Nucleosynthesis}

Supernova nucleosynthesis is commonly divided into several ``processes'', each of which is impacted by neutrino-nucleus interactions. (1) Explosive nucleosynthesis occurs as a result of compressional heating by the supernova shock wave as it passes through the stellar layers.  In the inner layers of the ejecta, where iron group nuclei result from $\alpha$-rich freezeout, interactions with neutrinos alter the neutronization, changing the ultimate composition.  (2) Neutrino nucleosynthesis or the ``$\nu$'' process occurs due to neutrino-induced nuclear transmutations in the outer stellar layers followed by shock heating. (3) The rapid neutron capture or ``r'' process may occur in the neutrino-driven wind that emanates from the proto-neutron star after the explosion is initiated. The neutrinos both drive the wind and interact with the nuclei in it.  Early phases of this wind have also been suggested as the source of light p-process nuclei \cite{HWFM96}.  Thus, neutrino-nucleus interactions are important to all core collapse supernova nucleosynthesis processes.

\subsection{Neutrinos and the $\alpha$-Rich Freezeout} 

One common property exhibited by recent spherically symmetric Boltzmann
simulations \cite{MLMH01,RaJa00} is a decrease in the neutronization (which is equivalent to an increase in $Y_e$) of the inner layers of the ejecta due to neutrino interactions.  This is a feature that current parameterized nucleosynthesis models can not replicate because they ignore the neutrino interactions.  The neutronization of the ejecta is important because galactic chemical calculations and the relative neutron-poverty of terrestrial iron and neighboring elements strongly limits the amount of neutronized material that may be ejected into the interstellar medium by core collapse supernovae \cite{Trim91}.  Those previous multidimensional models for core collapse supernovae that did produce explosions tended to greatly exceed these limits (see, \eg, \cite{JaMu96,HBHF94,KPJM00}).  To
compensate, modelers have been forced to posit the fallback of a
considerable amount of matter onto the neutron star, occurring on a
timescale longer than was simulated.  While the decreased
neutronization seen in Boltzmann models reduces the need to invoke
fallback, it also makes any fallback scenario more complicated, since the
most neutron-rich material may no longer be the innermost.  Because of the impact of the neutrinos on the nucleosynthesis, we fully expect the nucleosynthesis products from Boltzmann explosion simulations to be qualitatively different, both in composition and spatial distribution, from either parameterized bomb \cite{ThNH96} or piston \cite{WoWe95} nucleosynthesis models or the present generation of models of the core collapse mechanism.

\begin{figure}[t]
\begin{center}
\includegraphics[angle=-90,width=\textwidth]{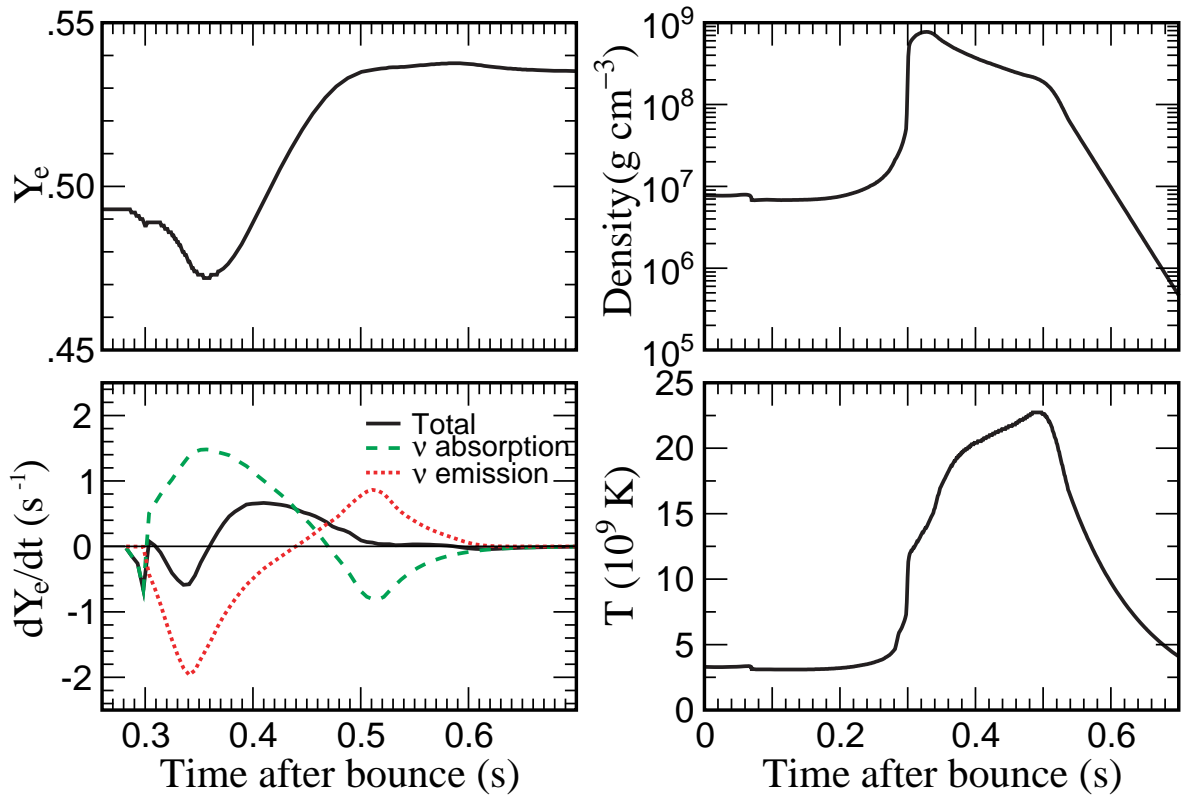}
\caption{The temporal history of the electron fraction, temperature and density for a zone in the innermost ejecta of a supernova model.  Note that the effects of neutrino capture dominate the effects of neutrino emission ($e^{\pm}$ capture), decreasing the neutronization.}
\label{fig:nunuc}
\end{center}
\end{figure}

Figure~\ref{fig:nunuc} demonstrates the effect of the neutrino interactions on the innermost ejecta.  In this example, the neutrino interactions of a model similar to \cite{MLMH01} have been modified in order to produce an explosion \cite{MHHL03}, but it is expected that realistic neutrino-driven explosions would be similar.  In the innermost ejecta, the shock fully dissociates the matter, so neutrino interactions with free nucleons dominate, producing a marked increase in the electron fraction.  In more distant regions, cooler peak temperatures will cause more poorly known $\nu$ and $\rm e^{\pm}$ interactions with heavy nuclei to dominate.  These interactions, as well as neutral current inelastic neutrino scattering off these nuclei \cite{BrHa91}, are also important to the thermal balance, affecting the $\alpha$-richness of the ejecta and, thereby, the abundance of important nuclei like \nuc{Ti}{44}, \nuc{Fe}{57}, \nuc{Ni}{58} and \nuc{Zn}{60} \cite{WoWe95}.  Thus, there is a clear need for improved neutrino nucleus interaction rates in order to accurately calculate the iron-peak nucleosynthesis from core collapse supernovae.  Because the degree of neutronization is much less than in deeper layers of the star, several nuclei of interest are directly accessible via the KARMEN technique: \nuc{Ca}{40}, \nuc{Sc}{45}, \nuc{V}{51}, \nuc{Mn}{55}, \nuc{Co}{59}.  However, theoretical calculations will still be necessary to provide full coverage of the many species present in significant concentrations.

\subsection{Neutrino Nucleosynthesis}

Neutrino nucleosynthesis is driven by the spallation of
protons, neutrons, and alpha particles from nuclei in the overlying
stellar layers by the intense neutrino flux that is emanating 
from the central proto-neutron star powering the supernova \cite{WHHH90}. 
Moreover, neutrino nucleosynthesis continues after the initial 
inelastic scattering reactions and the formation of their 
spallation products. The neutrons, protons, and alpha particles 
released continue the nucleosynthesis through further reactions 
with other abundant nuclei in the high-temperature supernova 
environment, generating new rare species. Neutrino nucleosynthesis 
occurs in two stages: (1) through the neutrino irradiation and 
nuclear reactions prior to shock arrival and (2) through the 
continuation of nuclear reactions induced by neutrinos as the 
stellar layers expand and cool.  The suggestion has been made \cite{WHHH90} that neutrino nucleosynthesis is responsible for the production of \nuc{B}{11}, \nuc{F}{19}, and two of Nature's rarest isotopes: \nuc{La}{138} and \nuc{Ta}{180} . 

Observations of the abundance of boron vary linearly with metallicity, implying that primary mechanisms that operate early in the history of our galaxy produce as much of these isotopes as secondary (quadratic) mechanisms that operate after the Galaxy has been enriched with metals.  The competing mechanism for boron formation, cosmic ray spallation, is a secondary process.  According to current models, neutrino nucleosynthesis in supernovae, which is a primary process, favors the production of \nuc{B}{11} over \nuc{B}{10}.  However laboratory calibration of the spallation channels producing these two isotopes is needed.  Used in conjunction with future HST observations discriminating between \nuc{B}{10} and \nuc{B}{11}, this measurement would be invaluable in resolving this controversy and supporting (or refuting) the suggestion that neutrino nucleosynthesis in supernovae is an important source of \nuc{B}{11} in the Galaxy \cite{HMWH99}. \nuc{B}{11} and \nuc{B}{10} are produced through the following spallation channels:
\begin{eqnarray*}
\mathrm{\nuc{C}{12} (\nu,\nu' p)  \nuc{B}{11} } & & \\
\mathrm{\nuc{C}{12} (\nu,\nu' n)  \nuc{C}{11} (e^{+}\nu) \nuc{B}{11} } & & \\
\mathrm{\nuc{C}{12} (\nu,\nu' d)  \nuc{B}{10} } & & \\
\mathrm{\nuc{C}{12} (\nu,\nu' pn) \nuc{B}{10}.} & &
\end{eqnarray*}

The final abundance of \nuc{F}{19} produced in a supernova can serve as 
a ``supernova thermometer.'' If the abundance of \nuc{F}{19} produced in 
the supernova is attributed to neutrino nucleosynthesis, the ratio of
[\nuc{F}{19}/\nuc{Ne}{20}]/[\nuc{F}{19}/\nuc{Ne}{20}]$_{\odot}$ (the denominator is the measured ratio in the Sun) is a measure of the mu and tau 
neutrinosphere temperatures \cite{Haxt99}.  \nuc{F}{19} is produced through the following spallation channels:
\begin{eqnarray*}
\mathrm{\nuc{Ne}{20} (\nu,\nu' n) \nuc{Ne}{19} (e^+ \nu_e) \nuc{F}{19}} & & \\
\mathrm{\nuc{Ne}{20} (\nu,\nu' p) \nuc{F}{19}.} & &
\end{eqnarray*}
Recent models \cite{HKHL03}, using improved neutrino nucleus reaction rates, show marked decreases in the production of \nuc{F}{19}.  

No obvious astrophysical site for the production of the rare isotopes \nuc{La}{138} and \nuc{Ta}{180} has been proposed. That they can be produced via neutrino nucleosynthesis in supernovae is compelling, and may be very important in that their existence, however rare, may be a fingerprint of the neutrino process.  If so, they potentially provide powerful diagnostics of the physics of the outer layers of the supernova.  \nuc{La}{138} and \nuc{Ta}{180} are produced through the following spallation channels:
\begin{eqnarray*}
\mathrm{\nuc{La}{139} (\nu,\nu' n) \nuc{La}{138} } & & \\
\mathrm{\nuc{Ta}{181} (\nu,\nu' n) \nuc{Ta}{180}.} & &
\end{eqnarray*}
Recent models \cite{HKHL03} imply significantly larger production of these isotopes, enhancing the possibility that these isotopes originate in supernovae. 

Experiments to measure the cross sections for all of these spallation channels are worthy of consideration for experimental determinations at stopped pion facilities.

\subsection{Nucleosynthesis in the Neutrino-Driven Wind}

The astrophysical r-process (rapid neutron capture process) is responsible for roughly half of the Solar System's supply of elements heavier than iron.  While the nuclear conditions necessary to produce the r-process are well established (see, \eg, \cite{KBTM93}), the astrophysical site remains uncertain. The leading candidate is the neutrino-driven wind emanating from the proto-neutron star after a core collapse supernova is initiated \cite{WWMH94}. However, other plausible sites have been suggested \cite{WhCH98,FrRT99}.  As the neutrino-driven wind expands rapidly and cools, the nuclear composition is dominated by $\alpha$-particles and free neutrons with a small concentration of iron group nuclei.  As temperatures continue to drop, charged particle reactions ``freeze out'' while neutron capture reactions continue on the ``seed'' heavy nuclei present at freeze-out.  Neutron capture $(n,\gamma)$ reactions are balanced by their inverse photodisintegration $(\gamma,n)$ reactions, establishing an equilibrium between the free neutrons and the nuclei in the wind.  Because of the high concentration of free nucleons, this $(n,\gamma)$-$(\gamma,n)$ equilibrium among isotopes of the same element produces nuclei that are quite neutron rich.  $\beta$ decays of nuclei with half lives that are short compared to the time scale for the r-process link these $(n,\gamma)$-$(\gamma,n)$ clusters, producing nuclei with higher Z and leading to the synthesis of heavier elements \cite{Meye94}.

Qian \etal\ \cite{QHLV97} have demonstrated that neutrino-induced reactions can significantly alter the r-process path and its yields in both the $(n,\gamma) \leftrightarrow (\gamma,n)$ equilibrium phase and the ``postprocessing phase'' that occurs once these reactions fall out of equilibrium. In the presence of a strong neutrino flux, $\nu_{\rm e}$-induced charged current reactions on the waiting point nuclei at the magic neutron numbers $N=50,82,126$ might compete with beta decays and speed up passage through these bottlenecks. Also, neutrinos can inelastically scatter on r-process nuclei via $\nu_{\rm e}$-induced charged-current reactions and $\nu$-induced neutral-current reactions, leaving the nuclei in excited states that subsequently decay via the emission of one or more neutrons. This processing may for example shift the abundance peak at $A=195$ to smaller mass. Extending this, Haxton \etal~\cite{HLQV97} pointed out that neutrino postprocessing effects would provide a fingerprint of a supernova r-process. Eight abundances are particularly sensitive to the neutrino postprocessing: $^{124}$Sn, $^{125}$Te, $^{126}$Te, $^{183}$W, $^{184}$W, $^{185}$Re, $^{186}$W, and $^{187}$Re.  Observed abundances of these elements are consistent with the
postprocessing of an r-process abundance pattern in a neutrino fluence consistent with current supernova models.  On a more pessimistic note, Meyer, McLaughlin, and Fuller \cite{MeMF98} have investigated the impact of neutrino-nucleus interactions on the r-process yields and have discovered that electron neutrino capture on free neutrons and heavy nuclei (in the presence of a strong enough neutrino flux) can actually hinder the r-process by driving the neutrino-driven wind proton rich, posing a severe challenge to theoretical models.  This push to lower neutronization makes the early phases of the neutrino-driven wind a candidate for production of the light p-process nuclei like \nuc{Se}{74}, \nuc{Kr}{78}, \nuc{Sr}{84} and \nuc{Mo}{92} \cite{HWFM96}.

During the r-process and subsequent postprocessing in the supernova neutrino fluence, neutrinos interact with extremely neutron-rich, radioactive nuclei. Thus, relevant direct neutrino-nucleus measurements cannot be made.  However, indirect measurements of charged- and neutral-current neutrino-nucleus interactions on heavy stable nuclei for $A>80$ would be invaluable as a gauge of the accuracy of theoretical predictions.

\section{\bf Supernova Neutrino Detection}

The nineteen neutrino events detected by IMB and Kamiokande from SN1987A confirmed the basic supernova paradigm---that core collapse supernovae mark the formation of a neutron star and release copious amounts of neutrinos---and signaled the birth of extra-Solar-System neutrino astronomy. For a Galactic supernova, thousands of events will be seen by Super-K and SNO, which for the first time will give us detailed neutrino ``light curves'' and bring us volumes of information about the deepest regions in the explosion. In turn, these light curves can be used to test and improve supernova models, thereby improving predictions about the explosion and resultant nucleosynthesis.  Moreover, from these detailed neutrino light curves and an understanding of the effects of neutrino oscillations, interesting insight could be gained about the density structure of the supernova progenitor.  Among the neutrino-nucleus interactions of relevance for supernova neutrino detection are neutrino interactions on \nuc{H}{2}, \nuc{O}{16}, \nuc{Fe}{56} and \nuc{Pb}{206,207,208}.

\subsection{Deuterium}

Neutrino experiments that use heavy water, like the Sudbury Neutrino Observatory (SNO), can detect supernova neutrinos via four main channels: 
\begin{eqnarray*}
\mathrm{e^{-}(\nu,\nu') e^{-}} & &\\
\mathrm{d (\nu, \nu n) p} & &\\
\mathrm{d (\nu_e,e^{-} p) p} & &\\
\mathrm{d (\bar{\nu}_e, e^{+} n) n} & & 
\end{eqnarray*}
Measurement of the reaction $\mathrm{d(\nu_e,e^{-} p) p}$, which is being considered to calibrate the reaction $\mathrm{p (p,e^{+} \nu_e) d}$ (part of the pp chain of reactions powering the Sun), would also provide a calibration for heavy water neutrino detectors.  Monte Carlo studies suggest that for the source brightness predicted for the ORNL SNS, two years of data in approximately thirty fiducial tons of D$_{2}$O would yield a cross section measurement with an accuracy of a few percent \cite{workshop}, which in turn will enable a more accurate interpretation of the SNO data from the next Galactic supernova. 

\subsection{Oxygen}

The charged-current reaction $\mathrm{\nuc{O}{16} (\nu_e,e^{-}) \nuc{F}{16}}$ is the principle channel for electron neutrino interactions for thermal sources in the range $T_{\nu_{\rm e}}\geq 4-5$ MeV, and its rate
exceeds that of neutrino-electron scattering by an order of magnitude 
for $T_{\nu_{\rm e}}\geq 7-9$ MeV \cite{Haxt87}.  Moreover, the electron 
angular distribution is strongly correlated with the electron neutrino 
energy, providing a way to measure the incident neutrino energy and, 
consequently, the electron neutrino spectra. By inference, one would 
then be able to measure, for example, the electron neutrinosphere 
temperature, providing an additional supernova thermometer \cite{workshop}. 

\begin{figure}
  \begin{center}
    \includegraphics[width=\textwidth]{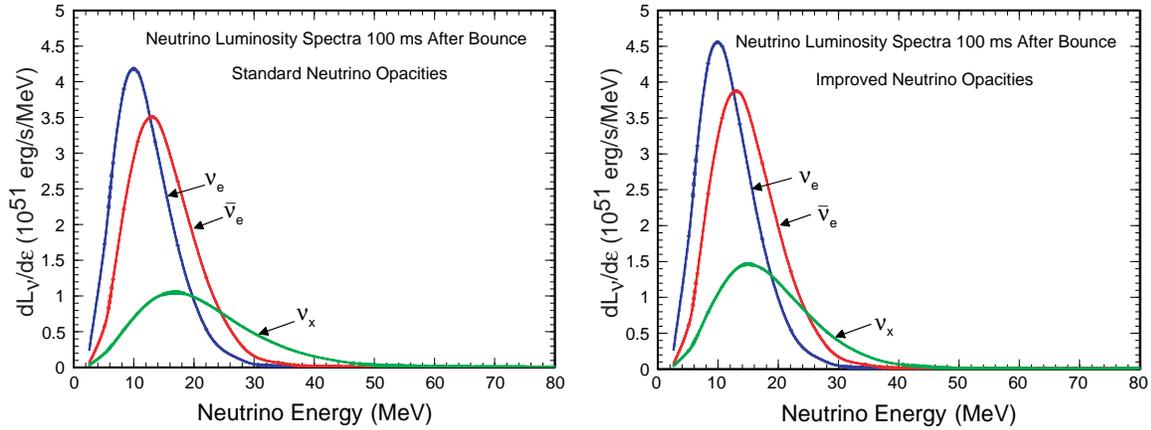} 
  \end{center}
  \caption{Comparison of the neutrino spectra from supernova simulations using the standard \cite{Brue85} opacities (left) and updated opacities (right; including the effects of neutino-nucleon absorption and elastic scattering \cite{RePL98}, neutrino-nucleon inelastic scattering and bremsstrahlung \cite{HaRa98}, and weak magnetism \cite{Horo02}). The simulations, initiated from a 13 M$_{\odot}$ progenitor, are fully general relativistic, and the spectra are computed at a radius of 500 km, 100 milliseconds after bounce \cite{Brue04}. }
   \label{fig:nspect}
\end{figure}

In addition, the appearance of back-angle electron emission from this reaction in, for example, Super-K would result from energetic electron neutrinos, more energetic than predicted by supernova models, providing further evidence for flavor oscillations and thereby information about the mu and tau neutrino spectra emanating from supernovae \cite{workshop}.  Mu and tau neutrinos in the stellar core couple to the core material only via neutral currents, whereas electron neutrinos and antineutrinos couple via both neutral and charged currents. As a result, the former decouple at higher density and, therefore, temperature, and have harder spectra.  Utilizing reactions on \nuc{O}{16}, Langanke, Vogel, and Kolbe \cite{LaVK96} have suggested a novel way of also unambiguously identifying mu and tau neutrino signatures in Super-K. The larger average energies for these neutrino flavors may be sufficient to excite giant resonances via the neutral-current reactions $\nuc{O}{16} (\nu_{\mu,\tau},\nu'_{\mu,\tau}) \nuc{O^{*}}{16}$. These resonances are above particle threshold and subsequently decay via the emission of protons, neutrons, and gamma rays. The gamma rays would provide the mu and tau neutrino signatures. The two decay channels are: $\nuc{O^{*}}{16} (,\gamma {\rm n}) \nuc{O}{15}$ and $\nuc{O^{*}}{16} (,\gamma {\rm p}) \nuc{N}{15}$.  However, potential channels for observing the mu and tau neutrinos from supernovae must be reexamined in light of recent work (see, \eg, \cite{ThBH00,Raff01,Brue04}), which indicates that nucleon-nucleon bremsstrahlung and the effects of nuclear recoil in neutrino-nucleon scattering significantly soften the mu and tau neutrino spectra, lessening their energy excess over electron neutrinos (see Figure \ref{fig:nspect}). 

Thus, accurate measurements of both charged- and neutral-current neutrino cross sections on \nuc{O}{16} would serve as a foundation for interpreting the neutrino data from the next Galactic core collapse supernova and for using the data to potentially observe the mu and tau neutrino spectra as it is emitted from the proto-neutron star.  An experiment to measure the cross section for:
\begin{equation*}
\mathrm{ \nuc{O}{16} (\nu_e,e^-) \nuc{F}{16}}
\end{equation*}
should be a high priority for a stopped pion facility.  Further useful experiments  could focus on the cross sections  for:
\begin{eqnarray*}
\mathrm{ \nuc{O}{16} (\nu_{\mu},\nu'_{\mu} n \gamma) \nuc{O}{15} }& & \\
\mathrm{ \nuc{O}{16} (\nu_{\mu},\nu'_{\mu} p \gamma) \nuc{N}{15} .}& &
\end{eqnarray*}

\subsection{Iron and Lead}

The use of iron and lead in supernova neutrino detectors like the proposed OMNIS detector would provide another way of detecting the mu and tau neutrino spectra in core collapse supernovae \cite{BoMu01}.  Iron has a sufficiently high threshold for neutron production via charged-current neutrino interactions that such production is 
negligible, whereas in lead neutrons are produced by both charged- 
and neutral-current interactions. Oscillations between the more 
energetic mu and tau neutrinos and the electron neutrinos would boost
the charged-current event rate while leaving the neutral-current 
rate roughly unchanged. Thus, the ratio of the event rate in lead 
to that in iron would serve as a further constraint on the extent of neutrino oscillations and the emitted mu/tau spectra.  However, this potential channel for observing the mu and tau neutrinos must also be reexamined in light of the softening of the mu and tau neutrino spectra.

To further the development of a detector like OMNIS, experiments to measure the neutrino-iron and neutrino-lead cross sections have been proposed. For iron, the neutral-current reaction:
\begin{equation*}
\mathrm{ \nuc{Fe}{56} (\nu,\nu' n) \nuc{Fe}{55}}
\end{equation*}
dominates. For lead, a total cross section would be measured 
resulting from the following neutral- and charged-current channels:
\begin{eqnarray*}
\mathrm{ \nuc{Pb}{208} (\nu,\nu' n) \nuc{Pb}{207}} & & \\
\mathrm{ \nuc{Pb}{208} (\nu,\nu' 2n) \nuc{Pb}{206}} & & \\
\mathrm{ \nuc{Pb}{208} (\nu_e,e^- n) \nuc{Bi}{207}}
\end{eqnarray*}
including also the channels for the isotopes \nuc{Pb}{206} and \nuc{Pb}{207}. For this reason, as well as their importance to nucleosynthesis, iron and lead cross section measurements should be among the first experiments at a stopped pion facility.

\begin{figure}
  \begin{center}
    \includegraphics[width=.8\textwidth]{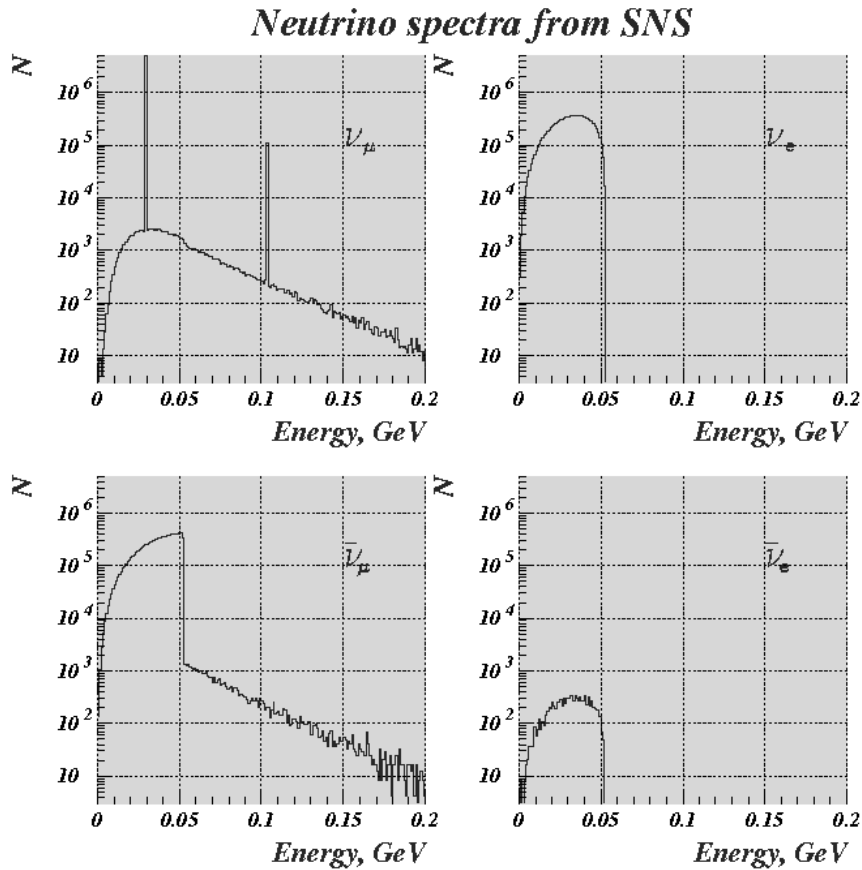} 
  \end{center}
  \caption{The expected neutrino spectra from the Spallation Neutron Source at ORNL.}
   \label{fig:snsspect}
\end{figure}

\section{Conclusion}
A new facility to measure neutrino-nucleus cross sections, such as those feasible at stopped pion facilities, would provide an experimental foundation for the many neutrino-nucleus weak interaction rates needed in supernova models.  With a neutrino source as intense as the ORNL SNS, for example, we are presented with a unique opportunity, given the overlap between the facility and supernova neutrino spectra (compare  Figure~\ref{fig:nspect}~\&~\ref{fig:snsspect}), to make such measurements. This would enable more realistic supernova models and allow us to cull fundamental physics from these models with greater confidence when their predictions are compared with detailed observations. 
Charged- and neutral-current neutrino interactions on nuclei in the stellar 
core play a central role in supernova dynamics and nucleosynthesis and are also important for supernova neutrino detection. Measurements of these reactions on select, judiciously chosen targets would provide an invaluable test of the complex theoretical models used to compute the neutrino-nucleus cross sections.

\section*{Acknowledgments}
The work has been partly supported by NASA under contract NAG5-8405, by the National Science Foundation under contract AST-9877130, by the Department of Energy, through the PECASE and Scientic Discovery through Advanced Computing Programs, and by funds from the Joint Institute for Heavy Ion Research. A.M. is supported by the Oak Ridge National Laboratory, which is managed by UT-Battelle, LLC, for the U.S. Department of Energy under contract DE-AC05-00OR22725.  The authors would like to acknowledge many illuminating discussions with Frank Avignone, John Beacom, Jeff Blackmon, Dick Boyd, David Dean, Yuri Efremenko, Jon Engel, George Fuller, Wick Haxton, Ken Lande, Karlheinz Langanke, Matthias Liebend\"{o}rfer,
Gabriel Martinez-Pinedo, Gail McLaughlin, Mike Strayer, and Friedel 
Thielemann, all of which contributed significantly to this manuscript.

\section*{References}
\providecommand{\bysame}{\leavevmode\hbox to3em{\hrulefill}\thinspace}

\end{document}